\documentclass[useAMS,usenatbib]{mn2e}
\usepackage{epsfig,graphicx,latexsym,amsmath,amssymb}
\usepackage{natbib}
\usepackage{hyperref}
\usepackage{mathrsfs}
\usepackage{lastpage}
\usepackage{bm}
\usepackage{verbatim}
\usepackage[labelformat=empty]{subfig}

\def\msun{{\,{\rm M}_\odot}}

\def\simlt{\lower.5ex\hbox{$\; \buildrel < \over \sim \;$}}
\def\simgt{\lower.5ex\hbox{$\; \buildrel > \over \sim \;$}}
\newcommand {\PN}{{\rm PN}}

\bibpunct[,]{(}{)}{;}{a}{}{,}

\title[Direct $N-$body simulations with spin corrections]
      {Relativistic mergers of compact binaries in clusters:\\ The fingerprint of the spin}

\author[P. Brem, P. Amaro-Seoane \& R. Spurzem] 
{Patrick Brem$^{1}$
                        \thanks{E-mail: Patrick.Brem@aei.mpg.de (PB)},
Pau Amaro-Seoane$^{1}$ \& Rainer Spurzem$^{2,\,3,\,4}$
   \\
$^{1}$Max Planck Institut f\"ur Gravitationsphysik
(Albert-Einstein-Institut), D-14476 Potsdam, Germany
\\
$^{2}$National Astronomical Observatories of China, Chinese Academy of
       Sciences, 20A Datun Lu, Chaoyang District, 100012, Beijing, China\\
$^{3}$Astronomisches Rechen-Institut, M{\"o}nchhofstra{\ss}e 12-14, 69120,
Zentrum f\"ur Astronomie, Universit\"at Heidelberg, Germany\\
$^{4}$Kavli Institute for Astronomy and Astrophysics, Peking
      University, China
}

\begin{document}

\pagerange{\pageref{firstpage}--\pageref{lastpage}} \pubyear{2012}

\maketitle

\label{firstpage}

\begin{abstract}
Dense stellar systems such as globular clusters and dense nuclear clusters are
the breeding ground of sources of gravitational waves for the advanced
detectors LIGO and Virgo. The stellar densities reached in these
systems lead to the dynamical formation of binaries at a rate superior to what
one can expect in regions of the galaxy with lower densities. Hence, these systems
deserve a close study to estimate rates and parameter distribution.
This is not an easy task, since the evolution of a dense stellar cluster
involves the integration of $N$ bodies with high resolution in time and space
and including hard binaries and their encounters and, in the case of
gravitational waves (GWs), one needs to take into account important
relativistic corrections.
In this work we present the first implementation of the effect of spin
in mergers in a direct-summation code, {\sc NBODY6}. We employ non-spinning
post-Newtonian corrections to the Newtonian accelerations up to 3.5
post-Newtonian (PN) order as well as the spin-orbit coupling up to
next-to-lowest order and the lowest order spin-spin coupling. We integrate spin
precession and add a consistent treatment of mergers.
We analyse the implementation by running a set of two-body experiments and then
we run a set of 500 simulations of a stellar cluster with a velocity dispersion set to a high
value to induce relativistic mergers to set a proving ground of the implementation.
In spite of the large number of mergers in our tests, the application of the
algorithm is robust. We find in particular the formation of a runaway BH
whose spin decays with the mass it wins, independently of the initial
value of the spins of the BHs. We compare the result with 500 Monte Carlo
realisations of the scenario and confirm the evolution observed with our
direct-summation integrator. More remarkably, the subset of compact objects
that do not undergo many mergers, and hence represent a more realistic system,
has a correlation between the final absolute spin and the initial choice for the
initial distribution, which could provide us with information about the evolution
of spins in dense clusters once the first detections have started.

\end{abstract}


\section{Introduction}

The field of GWs has reached a milestone in the last years with the build-up of
an international network of GW interferometers which have achieved their design
sensitivity.  The ground-based detectors LIGO and Virgo are undergoing major
technical upgrades that will increase the volume of the observable universe by
a factor of a thousand, which is referred to as the ``advanced''
configuration\footnote{\url{http://www.ligo.caltech.edu/advLIGO/},\\
\url{http://wwwcascina.virgo.infn.it/advirgo/}}.

Dense stellar systems such as globular clusters, galactic nuclei and, in
particular, dense nuclear clusters, are the breeding ground of the sources that
the advanced detectors can expect (see the recent updated review of
\citealt{BenacquistaDowning2011} and also \citealt{DowningEtAl2011}). More
remarkably, the event rate  of stellar-mass black hole binaries, the loudest
kind of source, will be likely {\em dominated by sources formed dynamically},
i.e. via stellar close interactions in these stellar systems
\citep{BanerjeeEtAl10,MillerLauburg09,DowningEtAl2010,BenacquistaDowning2011}.

The data that will be harvested from the advanced detectors will allow us to do
GW astrophysics. The construction of templates for matched filtering is crucial
in the searches for compact binaries. There have been efforts to construct
these templates by combining post-Newtonian calculations of the inspiral of the
binary with numerical relativity simulations of the merger and ringdown. Two
appealing approaches are the effective-one-body technique
\citep{BuonannoDamour1999,BuonannoEtAl2009} and the phenomenological hybrid
waveform modelling \citep{AjithEtAl07,AjithEtAl2009,SantamariaEtAl10}.

However, the search will be challenging for the simple reason that a
gravitational wave has not been detected yet. Reliable estimates of the
event rates for the different kinds of binaries and of the expected parameter
distribution will possibly be crucial for a successful detection. On the other
hand, once we have the data, we will be able to compare the observed rates and
parameters with the predictions derived from different models and thus
filter them. This will enlighten our understanding of the creation and
evolution of compact binaries in dense stellar systems.

The most accurate simulations of dense stellar clusters that we can do nowadays
are performed with the so-called ``direct-summation'' $N-$body algorithms. In
particular, the family of integrators of Sverre Aarseth has been in development
for many decades \citep{vonHoerner1960,vonHoerner1963,Aarseth1963}. Aarseth's {\sc Nbody6} includes
both {\em KS regularisation} (where KS stands for Kustaanheimo-Stiefel) and
{\em chain regularisation}: when particles are tightly bound or their
separation becomes too small, the system is regularised
\citep[see][]{KS65,Aarseth03} to avoid too small individual time steps and
numerical errors. It also employs the Ahmad-Cohen neighbour scheme
\citep{AhmadCohen73} and hierarchical, adaptive time steps.  We can hence
resolve and follow accurately individual orbits in the system.  In this article
we present the first modification of a direct-summation code, using {\sc
Nbody6}, that includes all non-spinning PN corrections up to 3.5PN order and
all spin contributions up to 2.5PN order, including spin precession equations.

\section{The formalism and its implementation}

\subsection{Correction of the accelerations}

We modify the acceleration computation as described in the pioneering work of
\cite{KupiEtAl06} (KAS06) to include relativistic corrections, which are based
on the post-Newtonian (PN) formalism for the interaction between two bodies. We
note that recently \cite{Aarseth2012} included an approximative implementation
for relativistic corrections in the new version of his code, {\tt NBODY7}. The
relative acceleration, in the center-of-mass form, including all PN corrections
used in the code can be written in the following way:

\begin{equation}
 \frac{d \vec v}{d t} = - \frac{G m}{r^2}[(1+A)\vec n + B \vec v] + \vec C_{\rm 1.5,SO} + \vec C_{\rm 2,SS} + \vec C_{\rm 2.5,SO},
\label{eq.acc}
\end{equation}

\noindent
where $\vec v = \vec v_1 - \vec v_2$ is the relative velocity vector, $m = m_1
+ m_2$ the total mass, $r$ the separation and $\vec n = \vec r/r$. $A$ and $B$ are coefficients that
can be found in \cite{BlanchetIyer03}. The spin terms $\vec C_{\rm N}$, where ${\rm N}$ denotes the PN order, are taken from \citep{FayeEtAl2006} and \citep{TagoshiEtAl01}.
SO stands for spin-orbit and SS for spin-spin coupling.

These corrections are valid for two {\em isolated} bodies and shall thus only
be applied to the Newtonian acceleration in the case of strong,
``relativistic'' pair-interactions where the perturbation by third bodies is
sufficiently small.  Because of this, we deem it reasonable to restrict the
implementation of PN terms to regularised KS pairs \citep[see][for
details]{KS65,Aarseth03}. For this reason we also choose the center-of-mass formulation shown in Eq. \ref{eq.acc} rather than the formulation in the general frame. These KS pairs are only formed when the interaction between
two bodies becomes strong enough so that the pair has to be regularised. During
the KS regularisation the relative motion of the companions is still far from
relativistic. Hence, only a small, relativistic subset of all regularised KS pairs
will need post-Newtonian corrections. In order to match
the order of accuracy of the KS integration in the code, we compute both the
acceleration as shown in Eq. \ref{eq.acc} as well as the analytical time
derivative.

To save computational costs we switch on the PN corrections only if one of the
following two conditions is fulfilled:

\begin{align}
 v & > \beta c \nonumber \\
 v & > \frac{\beta}{5} c,  {~\rm and~} \frac{g_\PN}{g} > \gamma_{\rm rel},
\label{eq.PNcond}
\end{align}

\noindent
where the parameters $\beta$ and $\gamma$ are chosen empirically to be $\beta=0.02$ and
$\gamma = 0.01$ and $g_\PN$ and $g$ are the PN acceleration and the Newtonian acceleration, respectively\footnote{In order to avoid confusion, we denote the acceleration with the letter $g$, the dimensionless spin parameter with $a$ and the semi-major axis with $\xi$.}. Note that this treatment differs from \cite{Aarseth2012}, who chooses
a staggered scheme to switch on first PN 2.5, and later PN1 or PN2. We always switch
on the complete set if equation~\ref{eq.PNcond} is fulfilled in order to maintain a correct orbit
integration under PN influence. The switch-on criterion for the PN terms does not depend on the Newtonian perturbation of the regularised pair. Thus we also apply PN corrections to binaries that are being influenced by a third body. However, we note that for strong perturbations, {\tt NBODY6} automatically breaks up the KS pair and uses a Chain regularization algorithm for more than two bodies, in which we do not include any PN treatment due to the complications that arise by having more than two dominant objects.

\subsection{Spin Precession}

In addition to the effects on the acceleration, the spin of compact objects
undergoes precession in relativistic two-body interactions. This is also taken
into account by integrating the spin precession equations

\begin{align}
 \frac{d \vec S}{d t} &= \frac{1}{c^2} \vec U_{\rm 1,SO} + \frac{1}{c^3} \vec U_{\rm 1.5,SS} + \frac{1}{c^4} \vec U_{\rm 2,SO} \\
 \frac{d \vec \Sigma}{d t} &= \frac{1}{c^2} \vec V_{\rm 1,SO} + \frac{1}{c^3} \vec V_{\rm 1.5,SS} + \frac{1}{c^4} \vec V_{\rm 2,SO} \\
 \vec S &= \vec S_1 + \vec S_2 \\
 \vec \Sigma &= m\left(\frac{\vec S_2}{m_2}-\frac{\vec S_1}{m_1}\right).
\end{align}
$\vec S$ and $\vec \Sigma$ describe the spin state of the pair.
The individual terms for $\vec U_N$ and $\vec V_N$, where $N$ denotes the PN order, can be found in \citep{FayeEtAl2006} and \citep{BuonannoEtAl03}.

\subsection{Relativistic mergers}
\label{sec.relmerg}

Since
relativistic binaries lose energy via the dissipative 2.5PN acceleration term,
we need to consistently add a relativistic ``merger recipe'' in the standard version of the
code. For the purposes of our study, we must address the following points:

\begin{enumerate}
 \item The criterion for two bodies to be transformed into one
 \item A dynamically correct treatment of the ``loss'' of
       one object from the simulation
 \item Computation of the spin of the BH that is formed after coalescence
       from the spins and orbital angular momentum of the BHs that participated in
       the coalescence
\end{enumerate}

Post-Newtonian theory can only be applied to the inspiral of the binary, but
not to the actual merger and ringdown.  We choose for up to 3.5PN order a
cut-off distance of $5\,R_S$, with $R_S = 2 G
(m_1 + m_2)/c^2$ the combined Schwarzschild radius \citep{YunesBerti2008}. For any instantaneous separation below this value, the pair is merged into one body.

On the other
hand, the newly formed compact object must have a mass and a velocity
vector consistent with the conservation of linear momentum. Also, since we
are treating spinning compact objects, all BHs must have an initial spin
vector. As we will see ahead, in section \ref{ch:toycode}, we use a fitting
formula at the last integration step before merging the bodies, i.e. at a separation of $5\,R_S$,
to assign a new spin value to the merged system following the prescription
of \cite{RezzollaEtAl2008}.

The work we present in this article should be envisaged as a first testing of
the algorithm with a ``stress test'': Our goal is the integration of a large
number of relativistic mergers in a stellar cluster. We achieve this, as we
will see later, by setting initially the cluster in a relativistic stage with
an extremely large central velocity dispersion. In order to maximise the number
of mergers, we neglect the recoil of coalescing pairs, since merging BHs with
a very large recoiling velocity could leave the system. However, a priori it is
straightforward to implement a recipe for the gravitational recoil by following
a similar fitting formula as in. e.g. the work of
\cite{PollnetEtAl2007,LoustoEtAl10}.

\section{Testing the implementation}
\label{ch:toycode}

In this section we test the implementation itself in a direct-summation code.
We present tests with a two-body integrator based on the same routines as {\sc
Nbody6}, but restricted to a simple, regularised two-body system.  This is
exactly the part of the modification in the integration that we aim at
implementing in {\sc Nbody6}, and hence is a perfect testing ground of our
algorithm.

In order to do so, we will compare our simple integrations with theoretical
approaches. In this regard, the formul{\ae} of \cite{Peters64} are useful for
testing the orbital decay in the simple non-spinning case. For spinning pairs
we will check the precession frequencies and conservation of the total angular
momentum.

\subsection{Non-spinning, merging relativistic binaries}
\label{sub:peters}

In this section we compare the results of our approximation with the
derivation of \cite{Peters64} of the evolution of the eccentricity
and semi-major axis of a binary which is decaying via the emission of
GWs. His derivations are based on Keplerian orbits and mimic the 2.5
dissipative term in the post-Newtonian expansion.

\begin{align}
 \Bigl\langle \frac{d\xi}{dt} \Bigr\rangle & = -\frac{64}{5} \frac{G^3 m_1 m_2 (m_1 + m_2)}{c^5 \xi^3 (1-e^2)^{7/2}} \Bigl(1+\frac{73}{24}e^2+\frac{37}{96}e^4\Bigr)\nonumber \\
\Bigl\langle \frac{de}{dt} \Bigr\rangle & = -\frac{304}{15} e\frac{G^3 m_1 m_2 (m_1 + m_2)}{c^5 \xi^4 (1-e^2)^{5/2}} \Bigl(1+\frac{121}{304}e^2 \Bigr)
\label{eq:Peters}
\end{align}

\noindent
In the last equations $\xi$ is the semi-major axis, $e$ the eccentricity, $t$ the time, $m_1$ and $m_2$
the mass of the first and second star in the binary, $G$ is the gravitational
constant and $c$ the speed of light.  In the case of a circular binary, as
shown in \cite{Peters64}, one can solve the differential equation for a binary
with companion masses $m_1$, $m_2$ and initial semi-major axis $\xi_0$:

\begin{equation}
 \xi(t) = (\xi_0^4 - 4 \beta t)^{1/4}
\end{equation}
where
\begin{equation}
 \beta = \frac{64}{5} \frac{G^3 m_1 m_2 (m_1 + m_2)}{c^5}.
\end{equation}
This yields a decay time of $T_c(\xi_0) = \xi_0^4/(4 \beta)$.

\begin{figure}
\begin{center}
          {\includegraphics[width=\columnwidth]{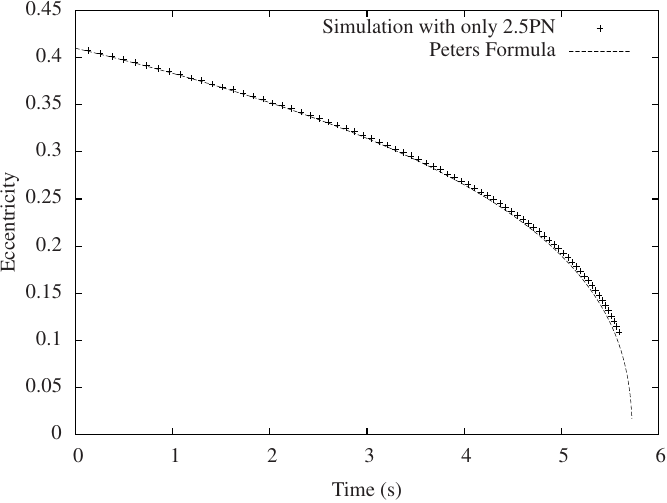}}
\caption
   {
Comparison of the eccentricity evolution of the two-body integration and Peters' approximation.
   }
\label{fig:petersonly25pnecc}
\end{center}
\end{figure}

\begin{figure}
\begin{center}
          {\includegraphics[width=\columnwidth]{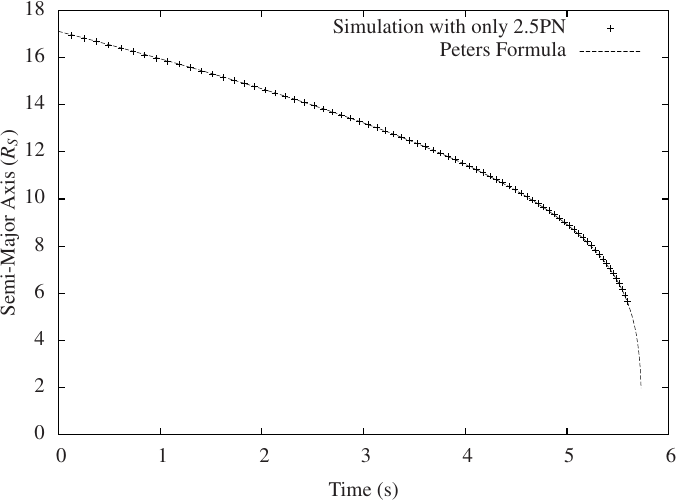}}
\caption
   {
Comparison of the semi-major axis evolution of the two-body integration and Peters' approximation.
   }
\label{fig:petersonly25pnsemi}
\end{center}
\end{figure}

In the general case of eccentric binaries, one can integrate Eq.
(\ref{eq:Peters}) numerically and compare the time evolution with the results
of our simulations. Since Peters' formula is only valid for the leading order
of gravitational radiation, we ``switch off'' the terms 1PN, 2PN, 3PN and 3.5PN
and only apply the 2.5PN correction.  In figure (\ref{fig:petersonly25pnecc})
and (\ref{fig:petersonly25pnsemi}) we show the time evolution of eccentricity
and semi-major axis for a system with two BHs of masses $m_1 = 10\msun$ and
$m_2 = 1\msun$. They agree very well up to the limit of validity of the
post-Newtonian expansion.

The 2.5 term only takes into account energy and angular momentum loss due to
GWs.  The 1 and 2 PN terms are conservative, they conserve energy and angular
momentum, and they are the main contribution to periapsis shift.

\begin{figure}
\begin{center}
          {\includegraphics[width=\columnwidth]{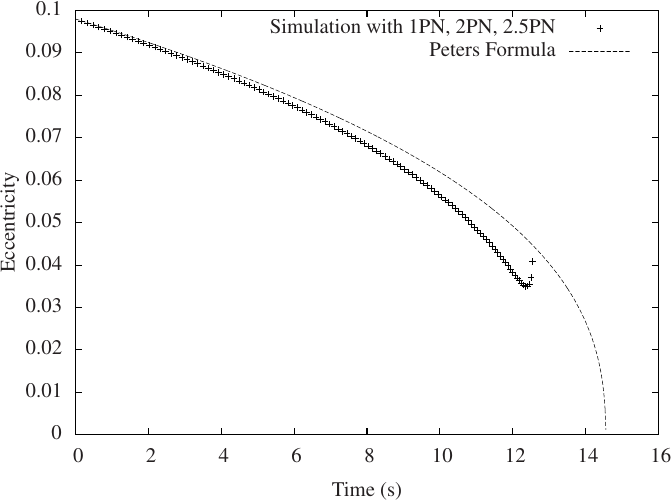}}
\caption
   {
Comparison of the eccentricity evolution of the two-body integration with 1PN, 2PN and 2.5PN terms and Peters' approximation.
   }
\label{fig:petersupto25pnecc}
\end{center}
\end{figure}

\begin{figure}
\begin{center}
          {\includegraphics[width=\columnwidth]{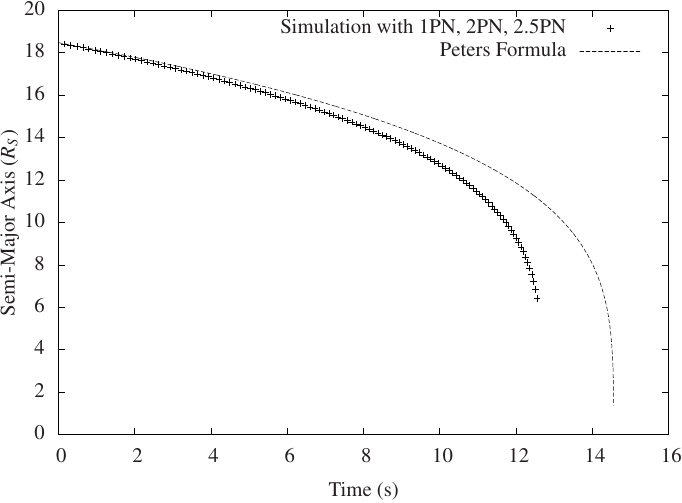}}
\caption
   {
Comparison of the semi-major axis evolution of the two-body integration with 1PN, 2PN and 2.5PN terms and Peters' approximation.
   }
\label{fig:petersupto25pnsemi}
\end{center}
\end{figure}

In figures (\ref{fig:petersupto25pnecc}) and (\ref{fig:petersupto25pnsemi}) we
show the time evolution for a binary in which we have taken into account the
correcting terms 1PN, 2PN and 2.5PN. Even though the 1 and 2PN terms are
conserving energy, the binary coalesces quicker than in the Peters
approximation, because they change the orbital velocity and thus the 2.5PN term
acts slightly stronger. The small rise in eccentricity very close to the merger is a known effect of the PN expansion at the limits of its region of validity.

The contribution at 3PN and 3.5PN order are small compared to the leading
order, but these terms cause the orbit to diverge when the binary enters the
last few $R_S$\footnote{Private communication with Seppo Mikkola and Cliff
Will}.  This is an important effect, since with PN terms up to order 2.5 one
could in principle let the system evolve until an overlap of the Schwarzschild
radii.  When including 3PN and 3.5PN, on the other hand, this becomes
impossible and in order to avoid unphysical, divergent behavior one has to
abort the integration at larger separations. For this reason we choose the
criterion $r = 5 R_S$ where $r$ is the instantaneous separation and $R_S$ is
the combined Schwarzschild radius.

\subsection{Spinning binaries}

\subsubsection{Precession of angular momenta}

In post-Newtonian theory, the Newtonian angular momentum $\vec L_N = \vec x
\times \vec p$, with $\vec p = \vec r \times m\,\vec v$, is no longer
conserved. In the case of non-spinning bodies, the direction of $\vec
L_N$ is conserved and only the modulus $L_N$ is gradually radiated away during
inspiral. However, in the case of spinning bodies this no longer holds
\citep{Kidder1995}. Nonetheless, as in electromagnetic theory, both the total
spin vector $\vec S$ and the angular momentum vector $\vec L$ precess around
the total angular momentum vector $\vec J = \vec L + \vec S$. The angular
momentum vector we use differs from the usual Newtonian definition:

\begin{equation}
 \vec L = \vec L_N + \vec L_{\rm 1PN} + \vec L_{\rm SO} + \vec L_{\rm 2PN}.
\label{eq:trueL}
\end{equation}

\noindent
With this definition, $\dot {\vec J} = 0$ up to 2PN order. The 2.5PN order,
however, introduces radiation loss.  \cite{Kidder1995} estimated the precession
frequency to lowest order, i.e. $\vec L = \vec L_N$. In the case of a single
spinning body with mass $m_s$ in a system with total mass $m$, the precession
frequency of both $\vec S$ and $\vec L_N$ is given by

\begin{equation}
 \omega_p = \frac{G |\vec J|}{2 c^2 r^3} \left(1+3 \frac{m}{m_s}\right).
\end{equation}

As an example, let us consider a system of a maximally spinning black hole of
mass $m_s = 10 ~ M_{\odot}$ and a non-spinning companion of mass $m_2 = 1 ~
M_{\odot}$. We set the system on a circular orbit in the x-y plane with radius
$10^8 ~ cm$ with the initial spin of $m_s$ in x-direction. This gives a total
initial angular momentum of

\begin{equation}
 |\vec J| = \sqrt{L_z(t = 0)^2 + S_{1,x}(t=0)^2} = 1.12 \times 10^{44} \frac{\rm kg~m^2}{\rm s}
\end{equation}

\noindent
and thus a precession frequency of $\omega_p = 0.18 ~ \rm Hz$. We use non-spinning PN terms up to 3.5PN order and Spin-Orbit coupling up to next-to-leading order.

\begin{figure}
\begin{center}
          {\includegraphics[width=\columnwidth]{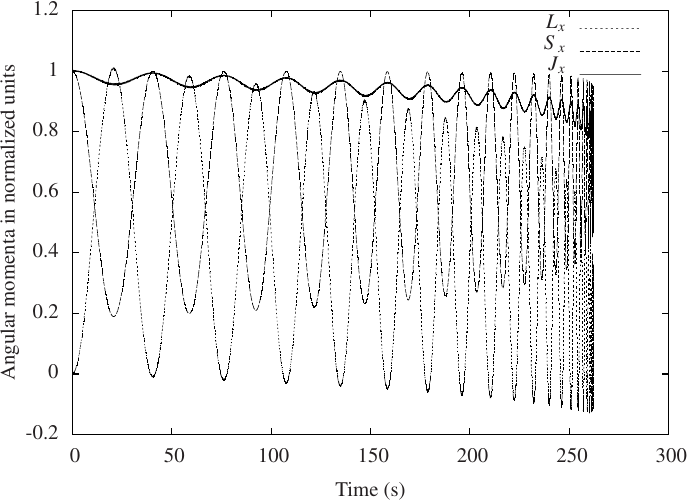}}
\caption
   {
Angular momentum precession in the case of one spinning body. The total Newtonian angular momentum vector $\vec J_N$ is approximately conserved.
   }
\label{fig:spin_precession}
\end{center}
\end{figure}

From figure (\ref{fig:spin_precession}) we can see that the approximate value
for the period of the first precession cycle is $(40.4 \pm 0.4)~ \rm s$. This
gives a value of $\omega_{p,\rm sim} = 0.15~ \rm Hz$. The small difference
comes from the fact that the calculation assumes the approximation $\vec L =
\vec L_N$, and we are already in a very relativistic regime.

\begin{figure}
\begin{center}
          {\includegraphics[width=\columnwidth]{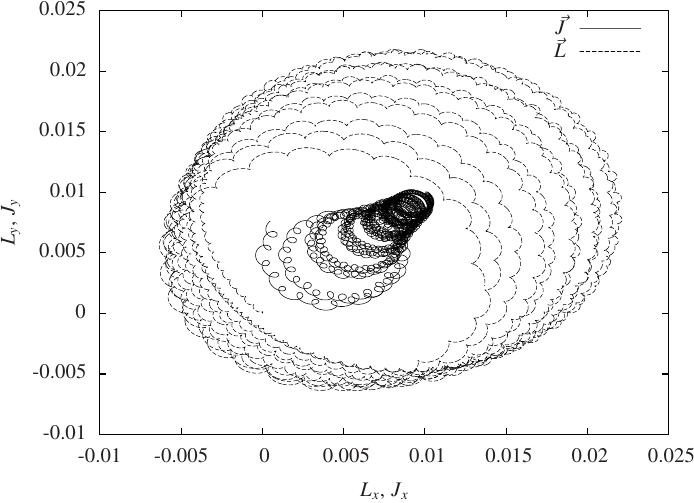}}
\caption
   {
X-Y-projection of the angular momentum precession in the case of two maximally spinning bodies. Both $\vec J_N$ and $\vec L_N$ are gradually radiated away as $\vec L_N$ precesses about $\vec J_N$.
   }
\label{fig:spin_precession_xy_proj}
\end{center}
\end{figure}

Even under the presence of spin-orbit precession, the direction of $\vec J_N$
should be conserved. Fig. (\ref{fig:spin_precession_xy_proj}) shows the
x-y-projection of $\vec J_N$ and $\vec L_N$ during an inspiral. One can see
that the direction of $\vec J_N$ is approximately constant but that the modulus
shrinks due to gravitational radiation.  During this process, $\vec L_N$
precesses about this direction. One can also see the wobbles in the precession
of the orbital plane given by $\vec L_N$, as described in the appendix of
\citep{Kidder1995}. This is due to the fact that in reality the corrected $\vec
L$ from Eq. (\ref{eq:trueL}) does the strict precession, which is not true for
the Newtonian value $\vec L_N$, and hence leads $\vec L_N$ to wobble about
the conserved $\vec L$.

The check of $\vec J$ conservation is a powerful way of testing the consistency
of the approach to estimate the spin and angular momentum in the code.

\subsection{Final spin approximation}

In our code we are subject to the limitations of our post-Newtonian approach,
which is not valid anymore when the relative speed becomes larger and larger,
i.e. a few Schwarzschild radii before the merger. For this, we adopt the
fitting formula of \cite{RezzollaEtAl2008}, derived
from numerical simulations that address in full general relativity the last
orbits of the binary, including merger and ringdown. We hence implement in the
code the following formula for the modulus of the final spin
\citep{RezzollaEtAl2008}

\begin{equation}
\begin{aligned}
  |\vec a_{\rm fin}| = &\frac{1}{(1+q)^2} \bigl[|\vec a_1|^2 + |\vec a_2|^2 q^4 + 2|\vec a_2| |\vec a_1| q^2 cos ~\alpha \\
 &+2(|\vec a_1| cos ~\beta + |\vec a_2| q^2 cos ~\gamma)|\vec l| q + |\vec l|^2 q^2 \bigr]^{1/2},
 \label{eq:rezzollaformula}
\end{aligned}
\end{equation}

\noindent
where $q = m_2/m_1$ is the mass ratio, $\vec a_1$ and $\vec a_2$ the dimensionless spin vectors and the angles are defined as

\begin{align}
cos ~\alpha & = \hat a_1 \cdot \hat a_2,\nonumber \\
cos ~\beta  & = \hat a_1 \cdot \hat l,\nonumber \\
cos ~\gamma & = \hat a_2 \cdot \hat l.
\label{eq:angles_spin}
\end{align}

\noindent
Therefore, so as to derive a value for the spin after merger, we need the
individual spin vectors $\vec a_1$, $\vec a_2$ and the orbital angular momentum
(OAM) at an arbitrary point in time during inspiral. $\vec l$ is a function of
the OAM, given by

\begin{align}
 |\vec l| = & \frac{s_4}{(1+q^2)^2} (|\vec a_1|^2 + |\vec a_2|^2 q^4 + 2|\vec a_1||\vec a_2| q^2 cos \alpha) \nonumber \\
 &+ \Bigl(\frac{s_5 \eta + t_0 + 2}{1+q^2} \Bigr) (|\vec a_1| cos \beta + |\vec a_2| q^2 cos \gamma) \nonumber \\
&+2\sqrt{3} + t_2 \eta +t_3 \eta^2,
 \label{eq:rezzollaformula2}
\end{align}

\noindent
where we use the fitting factors $s_i$, $t_i$ given in \cite{RezzollaEtAl2008}.
With Eq. (\ref{eq:rezzollaformula}) to (\ref{eq:rezzollaformula2}) in hand, one
can check whether in the regime in which PN is valid, the simulation is consistent with
this formula, in the sense that:

\begin{enumerate}

\item the total angular momentum must converge to the predicted absolute value

\item the predicted final value should be independent of the time until coalescence.

\end{enumerate}

\begin{figure}
\begin{center}
          {\includegraphics[width=\columnwidth]{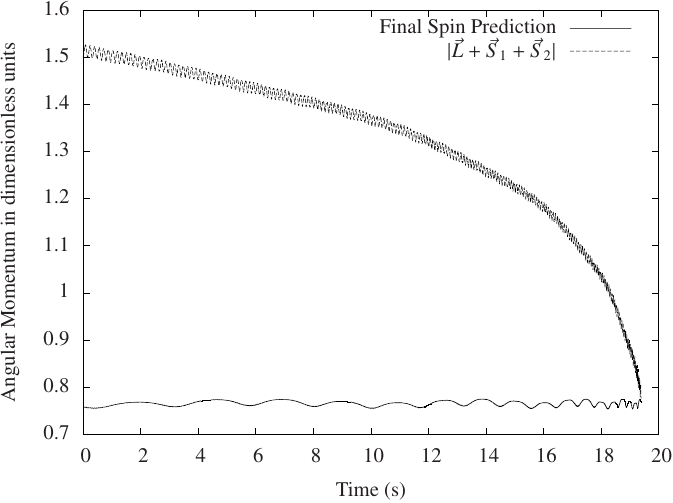}}
\caption
   {
Comparison between the current final spin prediction and the actual total angular momentum of the binary system.
   }
\label{fig:spinestimation}
\end{center}
\end{figure}

Figure (\ref{fig:spinestimation}) shows the time evolution of both the
predicted absolute value of the final spin at any given time during the
inspiral and the actual total angular momentum.  As one can see, for equal
masses this gives a consistent value. $\vec J$ is decreasing due to
gravitational radiation until it reaches the prediction. At the latest times close to the merger, there will remain a small difference between $\vec J$ and the predicted value due to the cut-off at $5\,R_S$ and due to other effects that are part of the numerical relativity simulations but not modelled in our PN integration.

\subsection{Energy conservation}

Since {\sc Nbody6} is a code to integrate Newtonian systems, it regularly
checks whether the total energy of the system is conserved within some
tolerance for numerical errors. In this work we have added relativistic terms
in the PN approximation, so that this is no longer the case: (i) The
dissipation, mainly by the 2.5PN term, causes a cumulative energy loss that has
to be tracked and subtracted from the total energy. On the other hand, (ii)
even the non-dissipative terms cause oscillations in the Newtonian energy,
since only the modified expression,

\begin{equation}
E = E_{\rm Newt} + E_{\rm 2.5PN,dis}+E_{\rm 1PN} + E_{\rm 2PN} + E_{\rm 3PN} + ...
\label{eq:energyconservation}
\end{equation}

\noindent
is conserved at any given time.  We thus calculate and subtract the corrections
up to 3PN order from the total energy in order to construct the conserved
quantity $E$. In this way we are able to verify energy conservation in the same
way as it is usually done in purely Newtonian codes. This works well if the
relativistic corrections are small. However, when $g_{\rm PN}/g \approx 1$ the
error induced by PN corrections will dominate and it becomes impossible to
verify the correct integration of the system. In order to avoid this, one could
decide an even larger distance threshold for merging two bodies into one or a
criterion based on the relative strength of the PN corrections.

\section{Stellar-mass binary mergers in a cluster: Sources of GWs for
ground-based detectors}

It is well-established that most galaxies should harbour a massive black hole
in their centre, with a mass of some $10^{\,6-9} M_{\odot}$ \citep[see
e.g.][]{FerrareseFord2005,FerrareseEtAl2001,KormendyGebhardt2001}.  The
densities observed may even exceed the core density of globular clusters by a
factor hundred, and hence achieve about $10^7-10^8 M_{\odot}~{\rm pc}^{-3}$.
Mass segregation creates a flow of compact objects towards the centre of the
system \citep{Lee1987,MEG00,KhalEtAl07,PretoAmaroSeoane10,Amaro-SeoanePreto11},
and may build up a cluster which could reproduce the effect of an MBH. Indeed,
this has been used as an alternative to explain phenomena related to cluster
evolution, like G1 and M15
\citep{BanerjeeKroupa11,GebhardtRichHo2002,BaumgardtEtAl2003a,BaumgardtEtAl2003b,
vanderMarelEtAl2002}. Nonetheless, for a globular cluster, compact objects such
as stellar black holes are very likely expulsed via three body interactions
\citep{PhinneySigurdsson1991,KulkarniHutMcMillan1993,SigurdssonHernquist1993,PortegiesZwartMcMillan00}.
\cite{Lee1995} proved that for $\sigma \gtrsim 300$ km/s the merger induced by
gravity loss in clusters with two components is shorter than the required
timescale for a third star to interact with a binary, so that clusters with
higher velocity dispersions will not run into that problem. In this section we
will test the robustness of our code by running simulations of dense stellar clusters
with a very high velocity dispersion to trigger a large number of relativistic
coalescences.

\subsection{Initial conditions}

To run a stress test on our implementation, we will consider that the clusters
are represented by an isotropic Plummer sphere containing $N = 1,000$ stellar
remnants of equal mass $m$. We use $N-$body units and choose a scaling
according to KAS06 to trigger a significant amount of relativistic mergers to
test the code. We set the central velocity dispersion to $\sigma_{\rm cen}
\approx 4,300\, {\rm km} {\rm s}^{-1}$, which is equivalent to fixing the ratio

\begin{equation}
  \frac{\sigma_{\rm cen}}{c} = \frac{1}{70}.
\end{equation}

\noindent
In other words, the speed of light ``in code units'' is $c = 70$.  We consider
therefore a cluster of compact objects with the same mass, spinning with a
dimensionless spin parameter $a$ and we consider three different initial spin
setups for the compact objects at the time $T=0$:

\begin{itemize}
  \item Non-spinning ($a = 0$)
  \item Maximally spinning in the z-direction ($a = 1$)
  \item Random magnitude and orientation
\end{itemize}

\subsection{Demonstration of a typical binary merger}

\begin{figure}
{\includegraphics[width=\columnwidth]{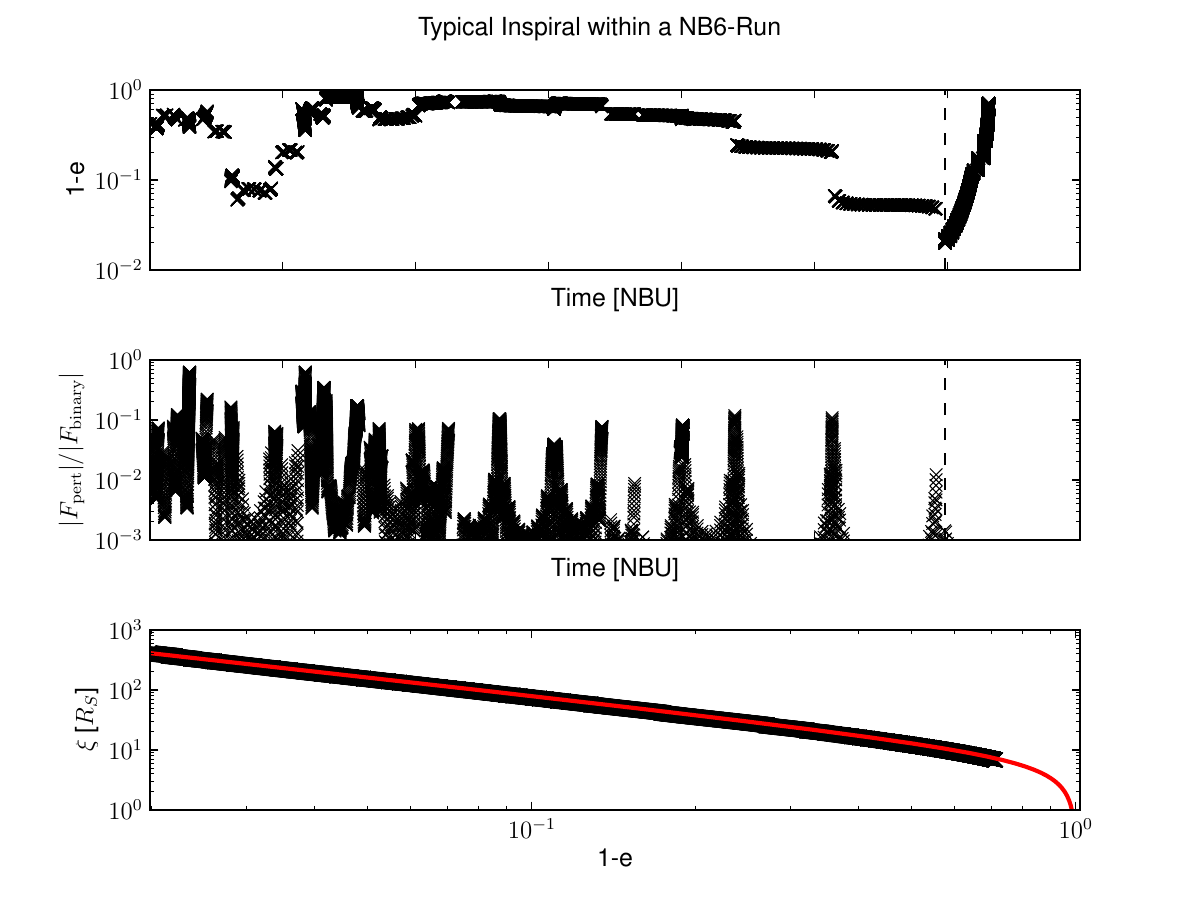}}
\caption
   {
     {\em Top panel:} Eccentricity evolution of one dynamically formed binary. First it is driven by Newtonian perturbations until the eccentricity reaches a critical value, from which the rapid circularization sets in. The dashed line marks the point from which we integrated Eq. (\ref{eq:Peters}) shown in the bottom panel. {\em Middle panel:} Perturbing force relative to the binary force. Strong changes in eccentricity are caused by strong Newtonian perturbations. {\em Bottom panel:} Inspiral as recorded in the simulation, compared to the analytical solution of Eq. (\ref{eq:Peters}) as the solid line.
   }
\label{fig.demonstration}
\end{figure}

We demonstrate here the evolution of a relativistic binary that has been formed dynamically within one of the non-spinning setups. Since we want to compare the decay to the approximation given by Eq. (\ref{eq:Peters}), only the dissipative 2.5PN term has been included. Figure \ref{fig.demonstration} shows the evolution of the orbital elements and the Newtonian perturbation by third bodies relative to the binary force. The eccentricity evolves due to Newtonian perturbation until it reaches a critical value and the GW driven inspiral sets in. From this point, the solution of Eq. (\ref{eq:Peters}) is plotted for comparison. We note that in all plotted data points, the PN terms have been switched on and we thus confirm the robustness of our implementation under the presence of strong Newtonian perturbations.

\subsection{Runaway growth}

\begin{figure}
{\includegraphics[width=\columnwidth]{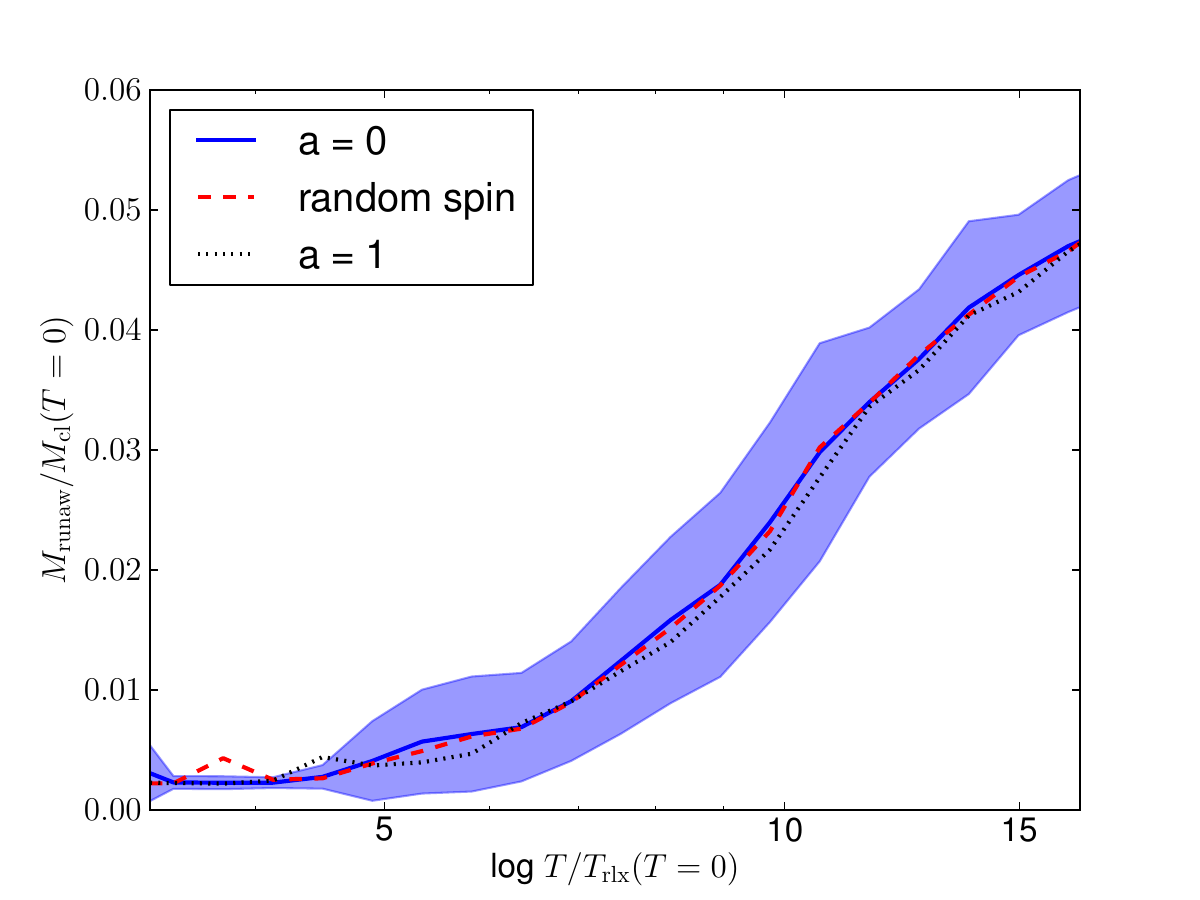}}
\caption
   {
Mass of the runaway body, $M_{\rm runaw}$, for each setup, averaged over 500 runs.
$M_{\rm cl}(T=0)$ is the total mass of the cluster at the time $T=0$
and $T_{\rm rlx}(T=0)$ the initial relaxation time of the cluster.
The
shaded area shows the standard deviation for the $a = 0$ case.
   }
\label{fig.runawaymass}

\end{figure}

\begin{figure}
{\includegraphics[width=\columnwidth]{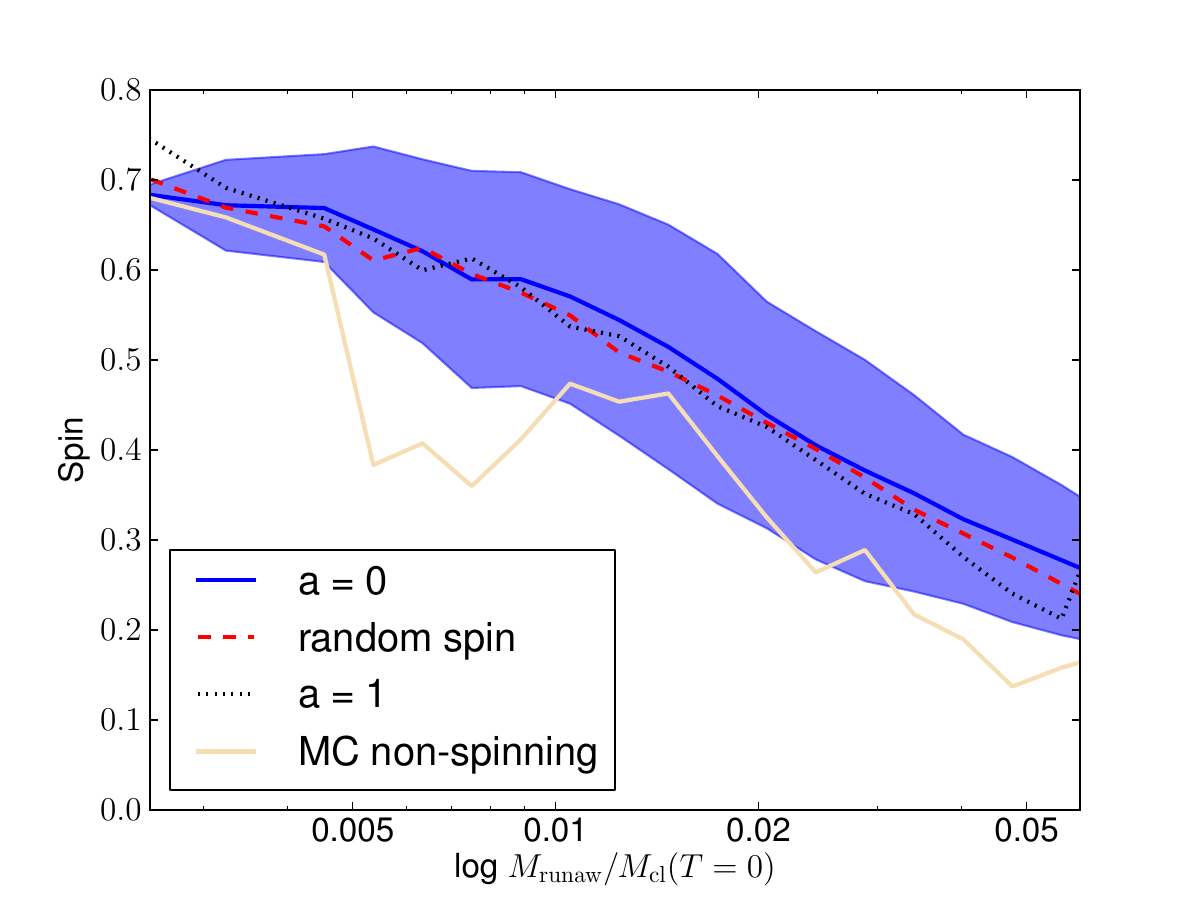}}
\caption
   {
Spin of the runaway body in each simulation, averaged over 500 runs. The shaded
area shows the standard deviation for the $a = 0$ case. All initial spin setups
lead to a similar evolution, except for the very first data point which is
slightly higher for the maximally spinning initial conditions.
   }
\label{fig.runawayspin}
\end{figure}

Because our system consists of very relativistic objects, almost any binary
that forms and is regularized will undergo a quick merger due to the loss of
orbital energy due to the dissipative 2.5PN term. Around the time of the core
collapse, i.e. after some $\sim 15\,T_{\rm rlx}(T=0)$, with $T_{\rm rlx}(T=0)$
the initial relaxation time of the cluster, a series of mergers leads to the
formation of one particular BH in the system that rapidly grows in mass and
becomes much more massive than the other objects. Therefore we say that the
object runs away in mass. This is a consequence of the increase in cross
section for GW capture. The time evolution of the mass of this runaway object
is shown in figure \ref{fig.runawaymass}. As we can see, after some $\sim
15\,T_{\rm rlx}(T=0)$, the runaway object has achieved $\sim 5\%\,M_{\rm
cl}(T=0)$, a value similar to the case studied in KAS06, their figure 1 around
450 time units.

\begin{figure}
\begin{center}
          {\includegraphics[width=\columnwidth]{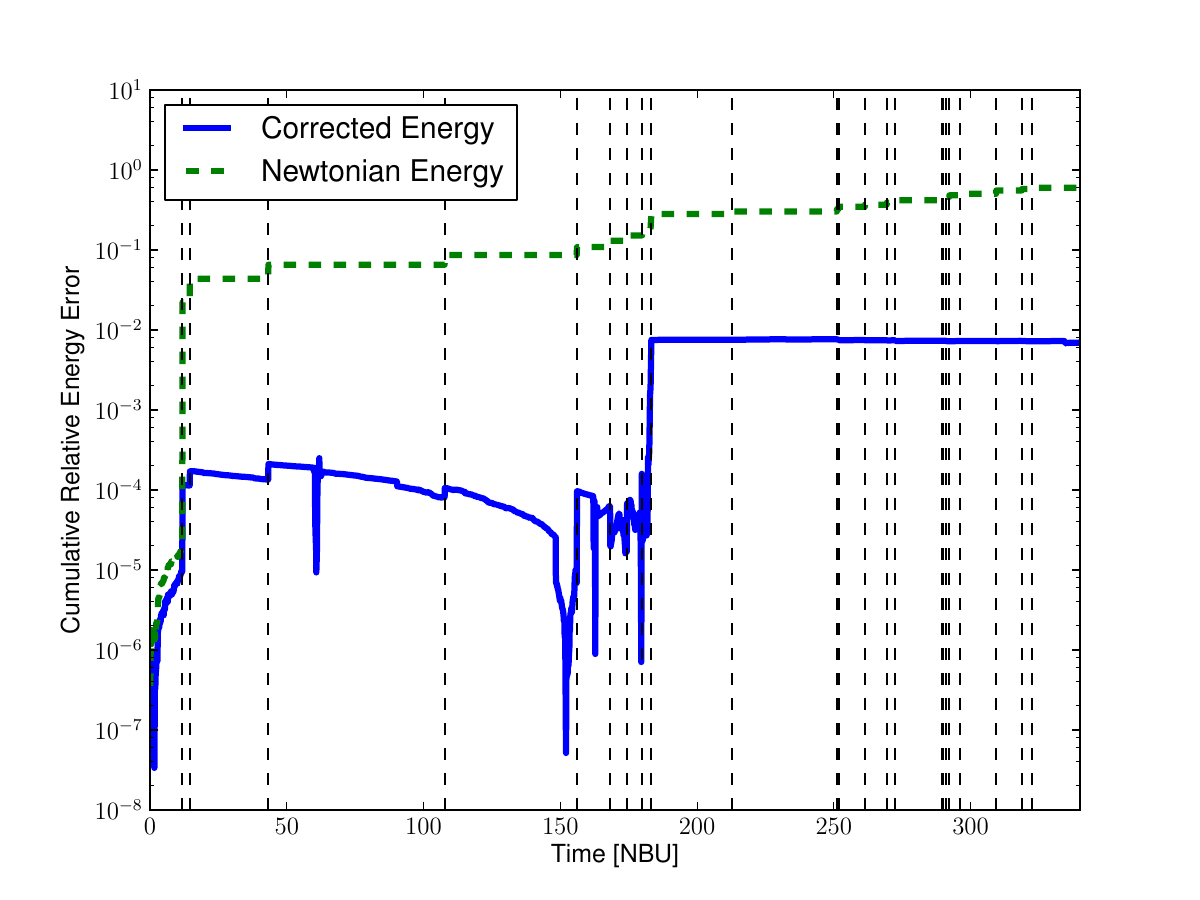}}
\caption
   {
Cumulative relative energy error in a typical simulation. In this case we have 22 mergers, indicated by the dashed vertical lines,
which cause the Newtonian energy error to grow significantly. Our alternative method to
check for energy conservation leads to smaller fluctuations.
   }
\label{fig:nbody_energy_error}
\end{center}
\end{figure}

An important issue that we need to address is the energy conservation in the
simulations. In figure \ref{fig:nbody_energy_error}
we show both the usual Newtonian energy and the corrected value, computed with Eq. (\ref{eq:energyconservation}) for a simulation with the same configuration as before but with $N = 2000$ bodies. The Newtonian energy error grows with every single merger due to the dissipative PN terms. The corrected value for the energy conservation in our approach fluctuates
significantly less and stays below $1\%$. The absolute value of the error depends on the nature of the merger: Head-on collisions dissipate the lowest amount of energy, while gradual inspirals lose the maximum amount before merger. The significant jump at $T = 183$ corresponds to a binary which has spent a very long inspiral time due to a low eccentricity and a high initial separation. This causes rather high errors in the numerical integration of the dissipated energy at 2.5PN order and thus contributes most to the total error, while some of the other mergers only cause relative errors of $\approx 10^{-4}$.

The absolute energy error depends crucially on the cut-off radius at which we end the integration and merge two bodies into one, because this sets the highest velocity we have to deal with in the binary. In this run we chose $10\,R_S$. For smaller values, even the corrected error grows to the order of the total energy of the system. We note that even with larger errors induced by the dissipative PN terms, the global behavior of the simulation is not affected by the particular choice of the merger radius. If one wants a poewrful energy conservation check it is reasonable to choose larger cut-off radii.

In order to be able to make a statistical comparison between each of the 3 spin
setups and the potential impact on the evolution of the runaway body, we
perform 500 simulations for each initial spin setup and show the mass averaged
over each time bin. We can see in figure \ref{fig.runawayspin} the evolution of
the spin for all 3 cases against the accumulated mass of the runaway object.
Its formation is approximately the same in all three different scenarios, and
consistent with the results of KAS06. Nonetheless, the precise point in time
where the onset of the runaway process takes places depends sensitively on the
scaling. In any case, the choice for the initial distribution of spins is
washed out and all three cases show a consistent evolution for the runaway
body.  We additionally perform 500 Monte Carlo realisations of the scenario
where one object merges with non-spinning compact objects coming in at random
angles using the same final spin prediction as in the $N-$body code, so that we
can test the statistical study. We depict the Monte Carlo spin evolution in
figure \ref{fig.runawayspin} and confirm that this evolution is consistent with
our $N-$body analysis within some scattering.

\subsection{Evolution of individual spins}

We now focus on the compact objects that have experienced only a few mergers.
While the evolution of the spin of the runaway object quickly washes out any
information regarding the initial spins, in the case of the other compact
objects that do not undergo so many mergers, there is a dependence on the
initial configuration even after core collapse. This is particularly
interesting, since a trend in the evolution of the spin measurable with the
advanced detectors would provide us with valuable information about the spin
evolution of compact objects in clusters.

As mentioned in section \ref{sec.relmerg}, we did not include BH recoil. 
For any BH merger involving significantly spinning BHs, the recoil velocity can exceed the
escape velocity and these merger products could thus leave the cluster. This means that the
distribution presented here contains BHs that might no longer be part of the cluster itself.

\begin{figure}
\begin{center}
          {\includegraphics[width=\columnwidth]{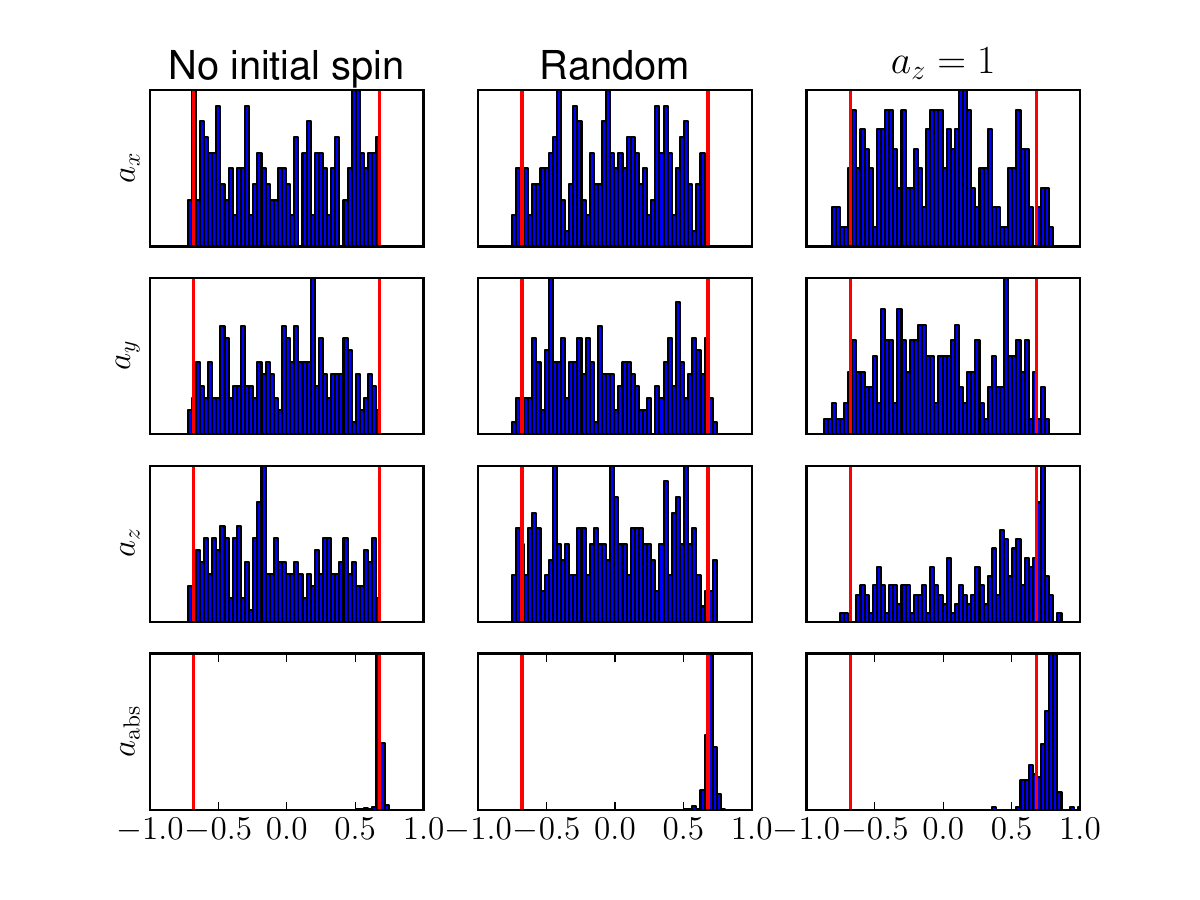}}
\caption
   {
Spin distribution for those objects that have undergone at least one merger
during the whole evolution of the cluster but not more than four. In the top panels
panels we show three different initial choices for the spin of the BHs. From
the left to the right we have first a cluster in which initially the BHs do
not have spin, then a random value and in the last column a maximally spinning
configuration around the $z$ direction. From the top to the bottom panels we
display the $x$-, $y$-, $z$- and absolute component of the spin ranging between
-1 and 1
($a_x,~a_y,~a_z,~a_{\rm abs}$ shown on the left y-axis of the panels, respectively).
The red lines depict the values -0.68 and +0.68.
   }
\label{fig:histspin}
\end{center}
\end{figure}

In figure~\ref{fig:histspin} we show the end distribution of spins for
different initial configurations of the spin distribution for an otherwise
identical system.   

The configuration which initially had no spins is useful for
comparison with the other systems. While the x-, y, and z-components
individually show no clear trend, the absolute value is $a_{\rm
abs} = (0.69 \pm 0.02)$. If we move on to the second configuration, in which we
initially assign all compact objects a spin but of random value, the final
distribution is scattered around the same value, displayed with a red line in
each of the panels at 0.68. In this case, the final value and standard
deviation are $a_{\rm abs} = 0.71 \pm0.03$. Finally, if we give all compact
objects initially a maximum value and set them in a preferred direction, which
we arbitrarily choose to be the positive z-direction, the final distribution has a value of
$a_{\rm abs} = 0.76 \pm 0.08$.

\begin{figure}
\begin{center}
          {\includegraphics[width=\columnwidth]{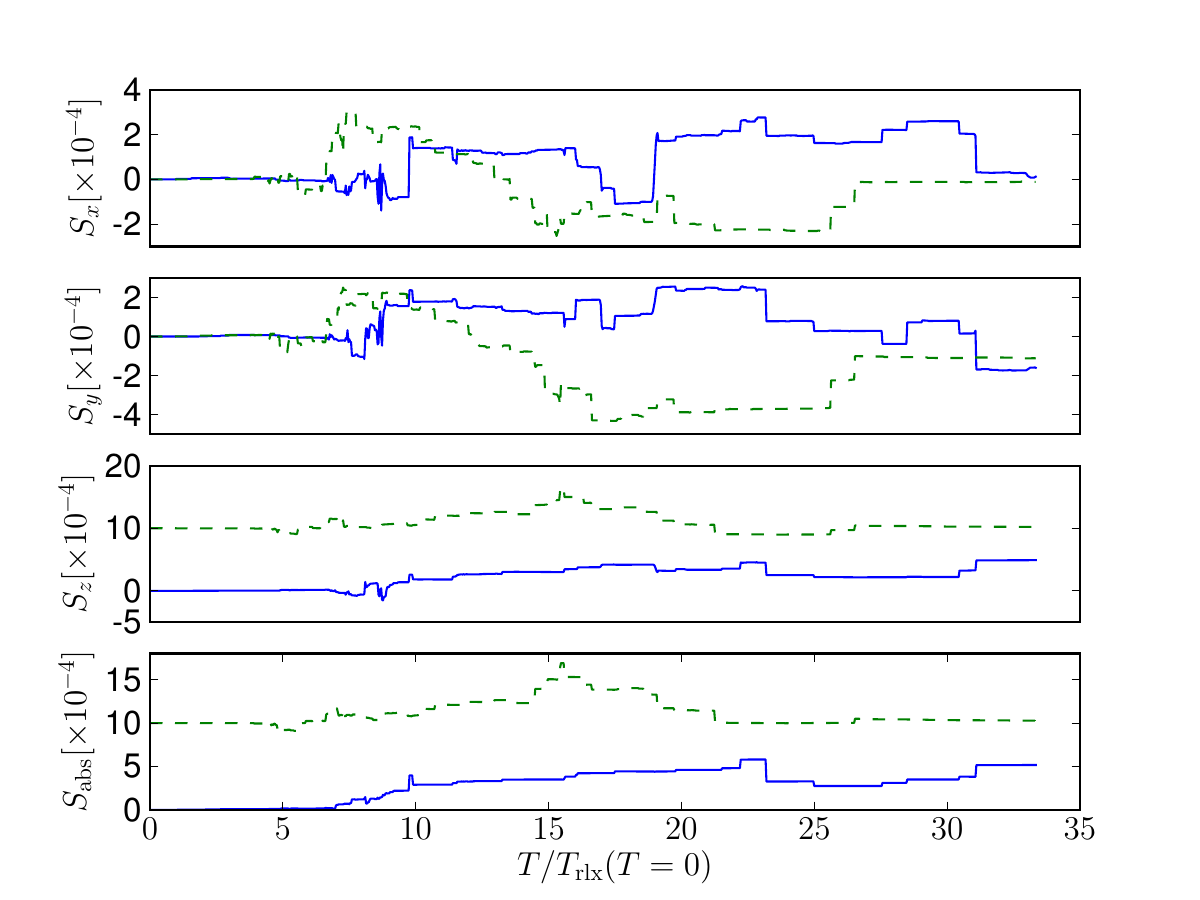}}
\caption
   {
{\em Dashed green:} Total spin angular momentum for a cluster in which the remnants are initially
maximally spinning in the z-direction. {\em Solid blue:} Total spin angular momentum for
an initially non-spinning configuration.
   }
\label{fig:totalspincomparison}
\end{center}
\end{figure}

In figure~\ref{fig:totalspincomparison} we can also see this dependence. In the
plot we display the time evolution of the total spin angular momentum in the
cluster, including the runaway object which carries most of the spin angular momentum.
In the case of an initially non-spinning configuration, the spin
builds up from orbital angular momentum and converges to a generic value in a
similar way to what we showed in figure~\ref{fig:histspin}. We are limited in
our analysis to derive the exact value to which the curve converges because of
an accumulation of numerical errors.

\section{Conclusions}

In this work we have presented the first implementation of the effect of the
spin for the treatment of relativistic mergers in a direct-summation $N-$body
integrator.  For that, we modify the calculation of the gravitational forces
among particles using post-Newtonian up to 3.5 post-Newtonian (PN) order and
the spin-orbit coupling up to next-to-lowest order and the lowest order
spin-spin coupling.

We then check our implementations by running a series of tests to compare with
results based on analytical derivations, for isolated two-body binaries and
confirm the robustness of our approach. We also present a way to check for the
correct integration of a system of $N$ particles based on tracking the
total energy, a usual test with this kind of integrators. Our method is valid
provided the number of relativistic mergers in the system is low.

The final acid test of the implementation is to compare the global dynamical
behaviour of a relatively large number of BHs with the new relativistic
behaviour for binaries with well-known results based on similar approaches.
More specifically, we run a similar test to that of KAS06 and obtain very
similar results, which confirms the correct incorporation of the new terms in
the code, since the initial spin distribution does not significantly change the
global evolution of the system. This is so, because if two non-spinning, equal
mass compact objects merge, the merger product will be spinning with $a \approx
0.68$ \citep{DamourNagar07} in the direction of the angular momentum.
Since in a Plummer sphere there is no preferred direction in the distribution
of the two-body angular momenta, this leads to a randomisation of the
non-spinning distribution quickly. In the scenario of two maximum spins in the
z-direction, i.e. individual spins of $S = G m^2/c$ with equal masses $m$, the
approximate angular momentum in the last stable orbit before merger is of the
same order and thus also rotate the spins and similarly wash out the initially
preferred direction.

For the larger subset of BHs that undergo a lower number of
coalescences, which is more interesting since it is closer to what one could
expect to see in a realistic cluster, we find that the evolution of the spin
for consecutive mergers has a trend that oscillates around the value predicted
by \cite{DamourNagar07}, but with a scatter that is a fingerprint of the
initial distribution of the isolated BHs, before they merged with any other
in the system. This is particularly interesting, since this trend is what will
determine the value of the spin that one can expect to see in globular
clusters, and should be carefully assessed when developing the waveform banks
to do the data analysis for the first detection.

Although the systems that we have explored in this work cannot be envisaged as
representative examples of the grounds for which we expect the advanced
detectors to observe relativistic mergers, the initial study of the behaviour
of the code is a requirement before we proceed to more realistic systems, and
has provided us with initial results which could play a crucial role in
detection.

In particular, an immediate goal of our next research will be the study of the
spin distribution and evolution in a dense stellar cluster with a realistic
number of stars and including stellar evolution and primordial binaries, such
as in \citep{DowningEtAl2010,DowningEtAl2011}, but with a more accurate direct
$N-$body integrator. The history and distribution of black holes in a dense
star cluster is also important for observing them in the electromagnetic
windows, since it determines e.g. number and distribution of X-ray binaries and
encounters between black holes and other compact objects such as neutron stars
or white dwarfs.

\cite{GierszEtAl2011} clearly show in their (non-relativistic) star cluster
simulations using the Monte Carlo code that quite a few BHs and BH-BH binaries
are formed and play a role for the dynamics of the central region. The presence
of BHs may explain the size differences between red and blue globular clusters
\citep{Downing2012} and affect the number of blue stragglers in a cluster
\citep{HypkiGiersz2013}. These papers also discuss that relativistic recoils
after merger not only important for the gravitational wave signal itself, but
it is an important ingredient for correct modelling of globular clusters.

The kind of analysis we have presented in this work will soon have interesting applications, taking into
account that the advanced ground-based detectors LIGO and VIRGO will have
achieved their desing sensitivity as soon as 2016-2017.

\section*{Acknowledgments}

We are indebted with Jonathan Downing and Peter Berczik for discussions.
PB and PAS thank the National Astronomical Observatories of China, the Chinese
Academy of Sciences and the Kavli Institute for Astronomy and Astrophysics in
Beijing, for an extended visit. This work has been supported by the Transregio
7 ``Gravitational Wave Astronomy'' financed by the Deutsche
Forschungsgemeinschaft DFG (German Research Foundation). It has also been
partially supported by the National Science Foundation under Grant No.
PHYS-1066293 and the hospitality of the Aspen Center for Physics.  RS
acknowledges support by the Chinese Academy of Sciences Visiting Professorship
for Senior International Scientists, Grant Number 2009S1-5 (The Silk Road
Project). The simulations were run on the ATLAS cluster of the
Albert-Einstein-Institute Hannover.

\label{lastpage}

\end{document}